\documentclass[reprint,superscriptaddress,preprintnumbers,nofootinbib,amsmath,amssymb,aps,prd]{revtex4-2}

\usepackage{graphicx}
\graphicspath{{figures/}}
\usepackage{dcolumn}
\usepackage{bm}
\usepackage{array}
\usepackage{amssymb}
\usepackage{color}
\usepackage{mathrsfs}
\usepackage{hyperref}
\hypersetup{colorlinks=true,linkcolor=blue,citecolor=blue,urlcolor=blue}
\usepackage{upgreek}
\usepackage{booktabs}
\usepackage{multirow}
\usepackage{threeparttable}
\usepackage{siunitx}
\usepackage{xcolor}
\usepackage{xurl}



\begin{document}

\title{Assessing Parameter Redundancy in Transformers for Jet Tagging}

\author{Huitong Cheng}
\affiliation{School of Physics and Electronics, Henan University, Kaifeng 475004, China}
\author{Yabo Dong}
\affiliation{School of Physics and Electronics, Henan University, Kaifeng 475004, China}
\author{Jun Fan}
\affiliation{School of Physics and Electronics, Henan University, Kaifeng 475004, China}
\author{Kun Wang}
\email[Corresponding author:]{kwang@usst.edu.cn} 
\affiliation{School of Physics, Faculty of Basic Sciences, University of Shanghai for Science and Technology, Shanghai 200093, China}
\author{Haijun Yang}
\affiliation{State Key Laboratory of Dark Matter Physics, Key Laboratory for Particle Astrophysics and Cosmology (MOE), Shanghai Key Laboratory for Particle Physics and Cosmology (SKLPPC),  School of Physics and Astronomy \mbox{\normalfont\&} Tsung-Dao Lee Institute, Shanghai Jiao Tong University, Shanghai 200240, China}
\author{Jingya Zhu}
\email[Corresponding author:]{zhujy@henu.edu.cn} 
\affiliation{School of Physics and Electronics, Henan University, Kaifeng 475004, China}
\author{Yifan Zhu}
\affiliation{State Key Laboratory of Dark Matter Physics, Key Laboratory for Particle Astrophysics and Cosmology (MOE), Shanghai Key Laboratory for Particle Physics and Cosmology (SKLPPC),  School of Physics and Astronomy \mbox{\normalfont\&} Tsung-Dao Lee Institute, Shanghai Jiao Tong University, Shanghai 200240, China}

\date{\today}

\begin{abstract}

Transformer-based jet taggers, such as the Particle Transformer (ParT) and the More-Interaction Particle Transformer (MIParT), achieve excellent discrimination by exploiting correlations among jet constituents, but often require more trainable parameters than earlier deep-learning taggers. 
In this paper, we investigate whether comparable discriminating power can be achieved with substantially fewer parameters. 
We introduce an hourglass structure that replaces the feed-forward networks (FFNs) in the attention blocks while leaving the particle-interaction attention unchanged. We also introduce a lightweight particle-embedding layer to replace the original dense embedding network. 
Applying both modifications to ParT and MIParT yields the hourglass (HG) variants ParT-HG and MIParT-HG, respectively.
We evaluate both models on benchmark datasets for top tagging and quark-gluon discrimination. 
Both variants retain comparable tagging performance, including background rejection at fixed signal efficiencies, while using only approximately 48\% and 39.7\% of the parameters of their respective baselines. 
On the larger JetClass dataset, accuracy and AUC decrease by less than 1\%, and background rejection also decreases for several signal classes. Overall, our approach provides an alternative way to reduce the parameter count of Transformer jet taggers while largely retaining their tagging performance.

\end{abstract}

\maketitle
\newpage


\section{Introduction}

\label{sec:intro}

Jet tagging identifies the origin of a jet from the particles observed inside it. It is important for precision measurements of Standard Model processes and for searches for new physics at high-energy colliders\cite{Larkoski:2017jix,Kogler:2018hem}. However, the origin of a jet cannot be observed directly. QCD radiation and hadronization produce a complex spray of particles that carries only indirect information about this origin. Observables such as the number of particles, jet mass, and jet charge, together with more detailed measures of jet substructure, provide useful information\cite{Thaler:2010tr,Larkoski:2013eya,Larkoski:2014gra,Dasgupta:2013ihk,Larkoski:2014wba,Krohn:2012fg}. These observables have clear physical meanings, but a selected set uses only part of the information contained in the measured particles.

Deep learning provides a way to use more of this information by taking the measured particles and energy deposits as inputs. Early work represented calorimeter energy deposits as jet images~\cite{Cogan:2014oua}, followed by convolutional-network classifiers operating on such images~\cite{deOliveira:2015xxd}. Later methods used the reconstructed particles themselves. Since their number varies from jet to jet and their ordering has no physical meaning, a jet can be treated as an unordered set\cite{Zaheer:2017wmg,Komiske:2018cqr}. ParticleNet uses a dynamic graph to learn local relations among particles\cite{Qu:2019gqs}, while Lorentz-equivariant networks build relativistic symmetry into the model\cite{Gong:2022lye,Bogatskiy:2023nnw,Brehmer:2024yqw}. Together, these approaches show that jet tagging can use both the properties of individual particles and the correlations among them.

Transformers extend this idea by allowing each particle to exchange information with all other particles through self-attention\cite{Vaswani:2017lxt}. ParT also supplies the attention layers with pairwise kinematic information, including angular separation, relative transverse momentum, energy fraction, and invariant mass\cite{Qu:2022mxj}. ParT achieves strong performance on JetClass and on the top tagging and quark--gluon datasets\cite{Qu:2022mxj,JetClass:2022zenodo}. MIParT gives the pairwise interactions a larger role and introduces More-Interaction Attention (MIA) in its early layers. It reports higher background rejection than ParT with fewer parameters\cite{Wu:2024thh}. These results motivate retaining the particle-interaction pathway when reducing the model size.

\begin{table*}[t]
\centering
\caption{Trainable-parameter distribution in baseline ParT and MIParT for the quark--gluon dataset ($C=13$ input features; two classes). P-MHA, MIA, and CA denote particle multi-head attention, More-Interaction Attention, and class attention, respectively. The indented FFN rows are included in the block totals immediately above. ``Others'' comprises the class token, final normalization, and classifier. ``All FFN sublayers'' sums the indented rows; ``Targeted modules'' additionally includes the particle embedding. Counts are rounded to $0.1$k, while fractions are calculated before rounding. For the top tagging dataset ($C=7$), only the particle-embedding count differs.}
\label{tab:baseline_parameter_concentration}
\small
\setlength{\tabcolsep}{6pt}
\newcommand{\ffnsubvalue}[1]{{\footnotesize\color{gray}\textit{(#1)}}}
\newcommand{\ffnsubdash}{{\footnotesize\color{gray}\textit{--}}}
\begin{tabular*}{\textwidth}{@{\extracolsep{\fill}}lrrrr@{}}
\toprule
& \multicolumn{2}{c}{ParT} & \multicolumn{2}{c}{MIParT} \\
\cmidrule(lr){2-3}\cmidrule(lr){4-5}
Module & Parameters (k) & Fraction & Parameters (k) & Fraction \\
\midrule
Particle embedding & 134.8 & 6.3\% & 102.0 & 15.5\% \\
Pair embedding & 9.6 & 0.4\% & 9.6 & 1.5\% \\
MIA blocks (including FFNs) & -- & -- & 191.1 & 29.0\% \\
\quad \textit{Included FFN contribution} & \ffnsubdash & \ffnsubdash & \ffnsubvalue{169.0} & \ffnsubvalue{25.7\%} \\
P-MHA blocks (including FFNs) & 1597.5 & 74.6\% & 253.5 & 38.5\% \\
\quad \textit{Included FFN contribution} & \ffnsubvalue{1065.0} & \ffnsubvalue{49.7\%} & \ffnsubvalue{169.0} & \ffnsubvalue{25.7\%} \\
CA blocks (including FFNs) & 399.4 & 18.6\% & 101.4 & 15.4\% \\
\quad \textit{Included FFN contribution} & \ffnsubvalue{266.2} & \ffnsubvalue{12.4\%} & \ffnsubvalue{67.6} & \ffnsubvalue{10.3\%} \\
Others & 0.6 & $<0.1\%$ & 0.3 & $<0.1\%$ \\
\midrule
\textbf{All FFN sublayers (sum)} & \textbf{1331.2} & \textbf{62.2\%} & \textbf{405.5} & \textbf{61.6\%} \\
Targeted modules (FFNs $+$ particle embedding) & 1466.0 & 68.4\% & 507.5 & 77.1\% \\
Total model & 2141.9 & 100.0\% & 657.9 & 100.0\% \\
\bottomrule
\end{tabular*}
\end{table*}

Several recent studies pursue smaller or faster Transformer jet taggers. IAFormer restructures interaction-aware attention to reduce the parameter count\cite{Esmail:2025kii}. Other methods use linear or hierarchical attention, omit explicit pairwise-interaction inputs, or combine model slimming and quantization\cite{Wang:2025dky,Zheng:2026juf,Petitjean:2025zjf,Wang:2026lqb}. Hardware-oriented studies further optimize Transformer taggers for low-latency FPGA inference\cite{Jiang:2024lvg,Laatu:2025xsw,Koski:2026rcc}. 
Quantization provides a complementary route to model compression by lowering the numerical precision of weights and activations. BitParT selectively replaces the linear projections in the ParT FFNs and classifier with 1-bit BitLinear layers, while retaining the attention, pairwise-interaction embedding, and input particle embedding in full precision~\cite{Rai:2025bitpart}. 
This raises a question: can the FFNs and particle embedding be redesigned to reduce the parameter count while retaining tagging performance and leaving the particle-interaction attention unchanged?

Each Transformer block contains attention and a feed-forward network (FFN). Attention exchanges information among particles; the FFN transforms each particle's feature channels separately\cite{Vaswani:2017lxt}. ParT and MIParT use the standard FFN shape $d\to4d\to d$, with a fourfold intermediate expansion\cite{Qu:2022mxj,Wu:2024thh}.

Table~\ref{tab:baseline_parameter_concentration} shows that the FFN sublayers alone contain approximately 62.2\% of the ParT parameters and 61.6\% of the MIParT parameters. The dense particle embedding contributes a further 6.3\% and 15.5\%, respectively. These distributions make the FFNs the primary target for parameter reduction, while the particle embedding provides another opportunity. Hourglass FFNs built from residual bottleneck subblocks offer a possible way to reduce the wide channel expansion\cite{Chen:2025hourglass,Liao:2026hourglass}. We therefore investigate whether the FFNs and particle embedding can be compressed while the particle-interaction attention remains unchanged.

We investigate this question in ParT and MIParT. We replace every standard FFN with stacked residual bottleneck subblocks and replace the dense particle embedding with grouped channelwise expansion followed by bottlenecked pointwise projections. The attention layers, pairwise-interaction representation, and their hyperparameters remain unchanged. This setup isolates the combined effect of the two architectural modifications from changes to attention. Applying both modifications gives the hourglass (HG) variants ParT-HG and MIParT-HG. We apply the same width-scaled design to MIParT-L to obtain MIParT-L-HG. We evaluate ParT-HG and MIParT-HG on the top tagging and quark--gluon datasets, and ParT-HG and MIParT-L-HG on JetClass using 10 million training jets. On the top tagging and quark--gluon datasets, the parameter counts decrease substantially with almost no change in tagging performance. On JetClass, accuracy and AUC decrease by less than 1\%, and background rejection also decreases for several signal classes.

Section~\ref{sec:method} describes the baseline architectures and the two parameter-reduction designs. Section~\ref{sec:results} presents results on the top tagging and quark--gluon datasets, the separate component ablations, and the larger-scale JetClass evaluation. Section~\ref{sec:conclusion} summarizes the conclusions.

\section{Model Architecture}

\label{sec:method}

\subsection{Jet Representation and Attention Mechanisms}
\label{subsec:baseline_architecture}

At the constituent level, a jet is specified by the particles it contains, and their ordering carries no physical information\cite{Zaheer:2017wmg,Komiske:2018cqr}. For constituent $i$, the feature vector $x_i$ contains $C$ particle-level observables, such as pseudorapidity $\eta$, azimuthal angle $\phi$, transverse momentum $p_T$, and energy $E$. Depending on the dataset, $x_i$ may also include particle-identification and track-displacement variables. ParT and MIParT therefore omit the positional encodings used for ordered sequences. The attention mechanism respects constituent exchange symmetry, so the jet classification is independent of particle ordering.

\begin{figure*}[t]
    \centering
    \includegraphics[width=\textwidth]{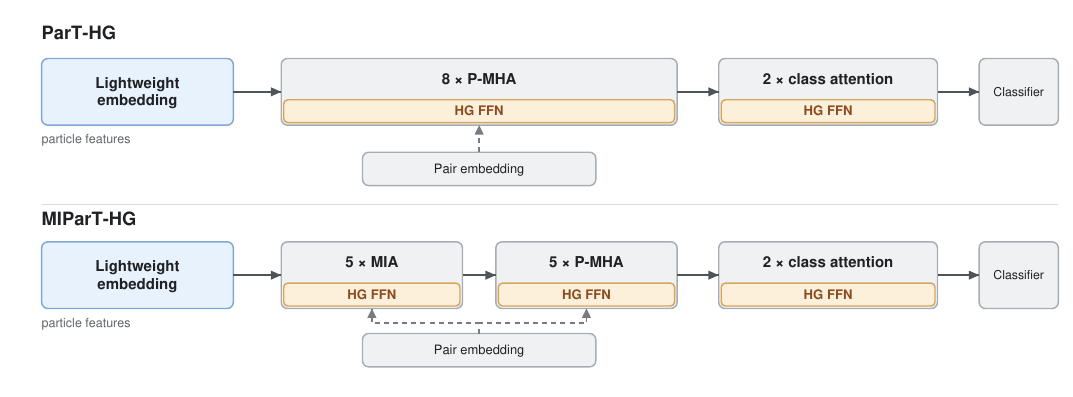}
    \caption{Overall architecture of the HG jet taggers. ParT-HG contains eight P-MHA blocks, whereas MIParT-HG contains five MIA blocks followed by five P-MHA blocks. Both models use two class-attention blocks. Blue and orange identify the replaced particle embedding and FFNs, respectively; gray identifies components retained from the corresponding baseline. The dashed lines show the unchanged pairwise-interaction pathway.}
    \label{fig:architecture_overview}
\end{figure*}

The four-momenta provide the second input at the particle-pair level. For each pair $(i,j)$, ParT and MIParT construct
\begin{equation}
  r_{ij}
  =
  \left(
  \ln\Delta_{ij},
  \ln k_{T,ij},
  \ln z_{ij},
  \ln m_{ij}^{2}
  \right).
  \label{eq:pair_features}
\end{equation}
where $\Delta_{ij}$ is the angular separation in rapidity and azimuth, and $k_{T,ij}=\min(p_{T,i},p_{T,j})\Delta_{ij}$. The remaining variables are $z_{ij}=\min(p_{T,i},p_{T,j})/(p_{T,i}+p_{T,j})$ and $m_{ij}^{2}=(p_i+p_j)^2$. Together, they encode the separation, relative transverse-momentum scale, momentum sharing, and invariant mass of the pair. A learned pair embedding maps $\{r_{ij}\}$ to the interaction representation $U$, which provides the pairwise input to the attention layers\cite{Qu:2022mxj,Wu:2024thh}.

In ParT, the pair embedding produces an eight-channel representation $U^{\mathrm{P}}$, with one channel for each attention head. Learned projections of the embedded particle features form $Q$, $K$, and $V$, and $U^{\mathrm{P}}$ is added to the scaled dot-product logits in particle multi-head attention (P-MHA):
\begin{equation}
  \mathrm{P\text{-}MHA}(Q,K,V;U^{\mathrm{P}})
  =
  \operatorname{Softmax}
  \left(
  \frac{QK^{T}}{\sqrt{d_{\mathrm{head}}}}
  + U^{\mathrm{P}}
  \right)V ,
  \label{eq:pmha}
\end{equation}
Here, $d_{\mathrm{head}}$ is the dimension of one attention head. P-MHA therefore combines correlations learned from the particle tokens with the explicit pairwise kinematics encoded in $U^{\mathrm{P}}$.

MIParT retains P-MHA but gives the pairwise interactions a more direct role in its early layers. A second pair embedding produces a 64-channel representation $U^{\mathrm{MI}}$, from which More-Interaction Attention (MIA) obtains the attention weights without query--key products:
\begin{equation}
  \mathrm{MIA}(U^{\mathrm{MI}},V)
  =
  \operatorname{Softmax}(U^{\mathrm{MI}})V .
  \label{eq:mia}
\end{equation}
The particle tokens enter through the value projection $V$, while $U^{\mathrm{MI}}$ determines how information is exchanged between particle pairs.

ParT contains eight P-MHA blocks, whereas MIParT applies five MIA blocks followed by five P-MHA blocks. Each attention operation is followed by a position-wise FFN. Both architectures then use two class-attention blocks to aggregate the particle representations into a class token, which is passed to the final classifier\cite{Qu:2022mxj,Wu:2024thh}. In the HG variants, we keep the pair embeddings, P-MHA, MIA, the block and head counts, and the class-attention mechanism unchanged. We replace only the particle embedding and the FFN sublayers, as summarized in Fig.~\ref{fig:architecture_overview}.

\subsection{Lightweight Particle Embedding}
\label{subsec:lightweight_embedding}

The particle embedding maps the measured features of each particle to a common $d$-dimensional token before information is exchanged among particles. The original dense embedding follows $C\to128\to512\to128$ in ParT and MIParT-L, and $C\to128\to512\to64$ in MIParT\cite{Qu:2022mxj,Wu:2024thh}. Its middle layer is therefore much wider than the final token dimension.

The original dense embedding mixes all particle-level observables from its first projection. We instead adapt the depthwise-separable design used in convolutional networks\cite{Chollet:2016xception,Sandler:2018mobilenetv2}. During the grouped $C\to4C$ expansion, each observable, such as $p_T$ or $\eta$, independently generates four intermediate features. The subsequent pointwise projections then combine features derived from different observables. The embedding therefore first processes each measured quantity separately and introduces cross-observable mixing only at a later stage, reducing the number of parameters required for the initial expansion. We use batch normalization in this convolutional embedding, with the statistics for each channel evaluated over the batch and constituent dimensions, while retaining LayerNorm in the Transformer subblocks.

\begin{figure}[t]
    \centering
    \includegraphics[width=\columnwidth]{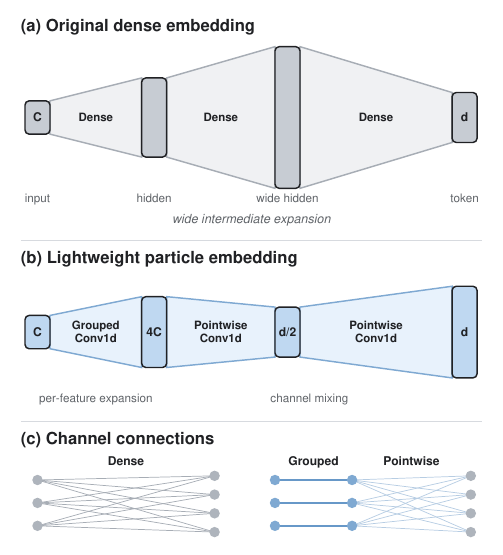}
    \caption{Comparison of the original dense and lightweight particle embeddings. Panels (a) and (b) show schematic channel-width profiles; the bar heights are not drawn to scale. The dense embedding contains a wide intermediate expansion, whereas the lightweight embedding uses a grouped Conv1d for per-feature expansion followed by two pointwise Conv1d layers for channel mixing. Panel (c) contrasts direct dense mixing with the grouped-then-pointwise channel connections.}
    \label{fig:lightweight_embedding}
\end{figure}

The lightweight embedding begins with batch normalization of the $C$ input features. A kernel-size-one grouped convolution with $C$ groups then maps $C\to4C$, so that each input feature is expanded independently. Two kernel-size-one pointwise convolutions subsequently implement $4C\to d/2\to d$ and mix information across feature channels. Each convolution is followed by batch normalization and a GELU activation. The complete mapping is therefore
\begin{equation}
  C
  \xrightarrow[\mathrm{groups}=C]{\mathrm{Conv1d}}
  4C
  \xrightarrow{\mathrm{Conv1d}}
  \frac{d}{2}
  \xrightarrow{\mathrm{Conv1d}}
  d .
  \label{eq:lightweight_embedding}
\end{equation}
All three convolutions act independently on each particle; correlations among different particles are still modeled by the unchanged attention layers. This separation of channelwise expansion and pointwise mixing follows the general structure of depthwise-separable and inverted-bottleneck networks\cite{Chollet:2016xception,Sandler:2018mobilenetv2}. Figure~\ref{fig:lightweight_embedding} compares the original and lightweight embeddings.

\begin{figure}[t]
    \centering
    \includegraphics[width=\columnwidth]{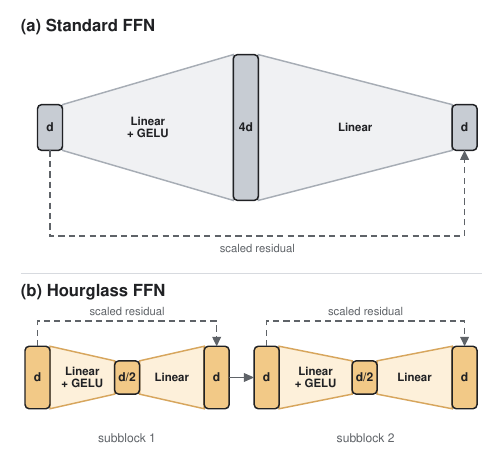}
    \caption{Comparison of the standard and hourglass FFNs; channel widths are schematic and are not drawn to scale. Panel (a) shows the standard $d\to4d\to d$ expansion with one scaled residual connection. Panel (b) shows the default hourglass configuration with two sequential $d\to d/2\to d$ bottleneck subblocks. Each subblock restores the representation to dimension $d$ and applies its own scaled residual connection before the next subblock.}
    \label{fig:hourglass_ffn}
\end{figure}

\subsection{The Hourglass Feed-Forward Network}
\label{subsec:hourglass_ffn}

Attention exchanges information among particles, whereas the FFN acts separately on the feature channels of each particle or class token. The standard FFN first expands the model dimension and then restores it, following $d\to4d\to d$\cite{Vaswani:2017lxt,Qu:2022mxj,Wu:2024thh}. Its two linear projections therefore contain $8d^2$ weights in total.

The conventional FFN maps each $d$-dimensional token to a $4d$-dimensional hidden space and then back to dimension $d$. The larger hidden dimension gives the network more intermediate dimensions in which to transform the information carried by each token, but the two projections also introduce many parameters. In fact, the FFN sublayers collectively account for most of the trainable parameters in both baselines. We therefore replace the $4d$ hidden layer with a narrower bottleneck of dimension $d_h<d$. Rather than using a single wide transformation, the hourglass FFN applies several $d\to d_h\to d$ subblocks in sequence, thereby exchanging width for depth. Each subblock makes a smaller update and has its own residual connection, which provides a direct path for information through the stack. This construction reduces the FFN parameter count without modifying the particle-interaction attention.

\begin{table*}[t]
\centering
\caption{Component-wise trainable-parameter counts for ParT and ParT-HG. The counts use the quark--gluon dataset with $C=13$ input features and a two-class output. P-MHA and CA denote particle multi-head attention and class attention, respectively. Component values are rounded independently to $0.1$k; percentage reductions are calculated before rounding.}
\label{tab:part_parameter_breakdown}
\small
\setlength{\tabcolsep}{6pt}
\begin{tabular*}{\textwidth}{@{\extracolsep{\fill}} l c r r r @{}}
\toprule
& & \multicolumn{2}{c}{Trainable parameters (k)} & \\
\cmidrule(lr){3-4}
Component & Multiplicity & ParT & ParT-HG & Reduction \\
\midrule
Particle embedding & 1  & 134.8  & 12.3  & 90.9\% \\
Pair embedding & 1  & 9.6  & 9.6  & 0.0\% \\
P-MHA sublayers, excluding FFNs & 8  & 532.5  & 532.5  & 0.0\% \\
CA sublayers, excluding FFNs & 2  & 133.1  & 133.1  & 0.0\% \\
FFN sublayers & 10 & 1331.2 & 341.8 & 74.3\% \\
Others & 1 & 0.6 & 0.6 & 0.0\% \\
\midrule
\textbf{Total} & & \textbf{2141.9} & \textbf{1030.0} & \textbf{51.9\%} \\
\bottomrule
\end{tabular*}
\end{table*}

\begin{table*}[t]
\centering
\caption{Component-wise trainable-parameter counts for MIParT and MIParT-HG. The counts use the quark--gluon dataset with $C=13$ input features and a two-class output. MIA denotes More-Interaction Attention. Component values are rounded independently to $0.1$k; percentage reductions are calculated before rounding.}
\label{tab:mipart_parameter_breakdown}
\small
\setlength{\tabcolsep}{6pt}
\begin{tabular*}{\textwidth}{@{\extracolsep{\fill}} l c r r r @{}}
\toprule
& & \multicolumn{2}{c}{Trainable parameters (k)} & \\
\cmidrule(lr){3-4}
Component & Multiplicity & MIParT & MIParT-HG & Reduction \\
\midrule
Particle embedding & 1  & 102.0 & 4.2   & 95.8\% \\
Pair embedding & 1  & 9.6   & 9.6   & 0.0\% \\
MIA sublayers, excluding FFNs & 5  & 22.1  & 22.1  & 0.0\% \\
P-MHA sublayers, excluding FFNs & 5  & 84.5  & 84.5  & 0.0\% \\
CA sublayers, excluding FFNs & 2  & 33.8  & 33.8  & 0.0\% \\
FFN sublayers & 12 & 405.5 & 106.8 & 73.7\% \\
Others & 1 & 0.3 & 0.3 & 0.0\% \\
\midrule
\textbf{Total} & & \textbf{657.9} & \textbf{261.3} & \textbf{60.3\%} \\
\bottomrule
\end{tabular*}
\end{table*}

We replace this wide intermediate representation with $N_{\mathrm{sub}}$ sequential residual bottleneck subblocks\cite{He:2015wrn,Chen:2025hourglass,Liao:2026hourglass}. For subblock $s$, the implemented transformation is
\begin{align}
  h^{(s)}
  &=
  \mathcal{D}_{a}
  \left[
  \sigma
  \left(
  W_{\downarrow}^{(s)}
  \operatorname{LN}^{(s)}(x^{(s)})
  + b_{\downarrow}^{(s)}
  \right)
  \right],
  \nonumber\\
  x^{(s+1)}
  &=
  \mathcal{D}_{o}
  \left[
  W_{\uparrow}^{(s)}
  \operatorname{LN}_{h}^{(s)}(h^{(s)})
  + b_{\uparrow}^{(s)}
  \right]
  +
  \gamma^{(s)}\odot x^{(s)} .
  \label{eq:hourglass_ffn}
\end{align}
Here, $x^{(s)}$ is the subblock input and $h^{(s)}$ is its bottleneck activation. The matrices $W_{\downarrow}^{(s)}$ and $W_{\uparrow}^{(s)}$ map $d\to d_h$ and $d_h\to d$, respectively, with corresponding biases $b_{\downarrow}^{(s)}$ and $b_{\uparrow}^{(s)}$. The function $\sigma$ is GELU, while $\mathcal{D}_{a}$ and $\mathcal{D}_{o}$ denote the activation and output dropout operations. The normalizations $\operatorname{LN}^{(s)}$ and $\operatorname{LN}_{h}^{(s)}$ act on the $d$ and $d_h$ channels, respectively, and $\odot$ denotes elementwise multiplication. The trainable vector $\gamma^{(s)}$, initialized to one, scales the skip connection channel by channel. Thus, each subblock returns to dimension $d$ before the next subblock is applied.

The P-MHA and MIA blocks use dropout rates of 0.1, while the class-attention blocks retain the zero-dropout setting of the baselines. Unless otherwise stated, we use
\begin{equation}
  N_{\mathrm{sub}}=2,
  \qquad
  d_h=\frac{d}{2}.
  \label{eq:hourglass_default}
\end{equation}
The leading projection-weight count is then
\begin{equation}
  N_{\mathrm{proj}}^{\mathrm{HG}}
  =
  2N_{\mathrm{sub}}dd_h
  =
  2d^2,
  \label{eq:hourglass_params}
\end{equation}
which is one quarter of the $8d^2$ term in the standard FFN. Biases, normalization parameters, and residual-scaling vectors are included in the complete parameter counts. The hourglass FFN is applied to every MIA, P-MHA, and class-attention block. Figure~\ref{fig:hourglass_ffn} shows the standard and hourglass structures.

\subsection{Compressed Model Configurations}
\label{subsec:model_config}

We apply the lightweight particle embedding and hourglass FFN jointly to the established ParT and MIParT architectures. For brevity, we append the suffix ``HG'' to models containing both modifications; although the suffix refers to the hourglass structure, it labels the complete compressed configuration in this work. ParT-HG retains the $d=128$ ParT backbone and uses the embedding path $C\to4C\to64\to128$, whereas MIParT-HG retains the $d=64$ MIParT backbone and uses $C\to4C\to32\to64$. In both models, each hourglass FFN uses the two residual bottleneck subblocks described above, with $d_h/d=1/2$ for each subblock. All other model components remain unchanged from their corresponding baselines.

Table~\ref{tab:part_parameter_breakdown} compares the trainable-parameter counts of ParT and ParT-HG. The lightweight particle embedding reduces its contribution from 134.8k to 12.3k parameters, while the hourglass FFNs reduce the combined FFN contribution from 1331.2k to 341.8k. The total parameter count therefore decreases from 2.14M to 1.03M, corresponding to a reduction of 51.9\%.

\begin{figure*}[t]
    \centering
    \includegraphics[width=\textwidth]{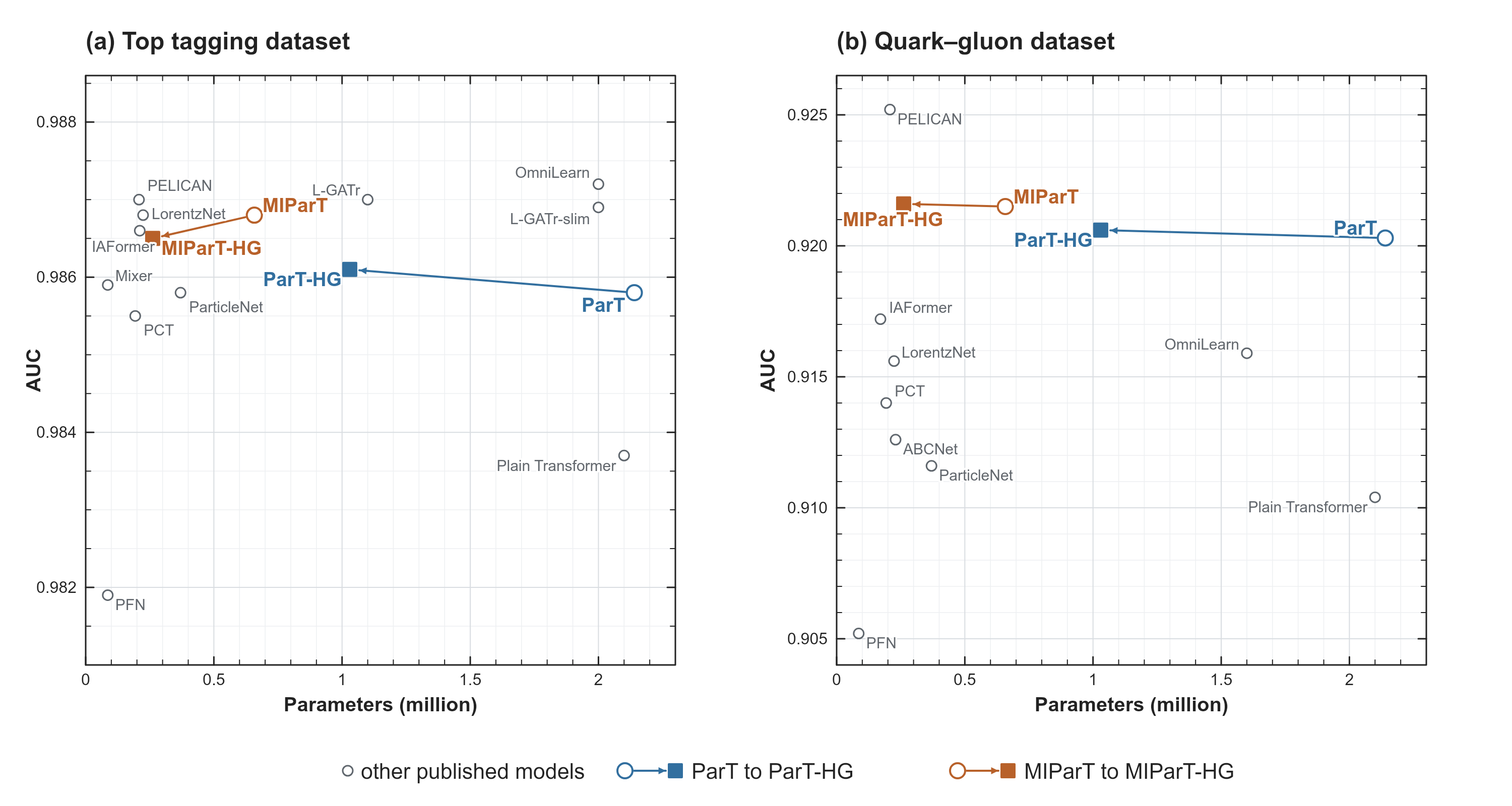}
\caption{AUC versus trainable parameter count for (a) the top tagging dataset and (b) the quark--gluon dataset. Arrows connect ParT to ParT-HG and MIParT to MIParT-HG, corresponding to parameter reductions of 51.9\% and 60.3\%, respectively.}
    \label{fig:auc_vs_parameters}
\end{figure*}

\begin{table*}[!t]
\centering
\small
\setlength{\tabcolsep}{6pt}
\caption{Performance of ParT, ParT-HG, MIParT, and MIParT-HG on the top tagging dataset and the quark--gluon dataset. Background rejection is reported at signal efficiencies of 50\% and 30\%. Parentheses in the last column give the parameter count relative to the corresponding baseline. 
}
\label{tab:binary_performance}
\begin{tabular*}{\textwidth}{@{\extracolsep{\fill}} l c c c c c @{}}
\toprule
Model & Accuracy & AUC & $\mathrm{Rej}_{50}$ & $\mathrm{Rej}_{30}$ & Parameters (fraction) \\
\midrule
\multicolumn{6}{l}{\textit{Top tagging dataset}} \\
\addlinespace[2pt]
ParT\cite{Qu:2022mxj} & 0.940 & 0.9858 & $413\pm16$ & $1602\pm81$ & 2.14M (100\%) \\
ParT-HG (this work) & 0.941 & 0.9861 & $450\pm1$ & $1638\pm3$ & 1.03M (48.1\%) \\
MIParT\cite{Wu:2024thh} & 0.942 & 0.9868 & $505\pm8$ & $2010\pm97$ & 657.9k (100\%) \\
MIParT-HG (this work) & 0.942 & 0.9865 & $483\pm13$ & $1861\pm26$ & 261.3k (39.7\%) \\
\addlinespace[5pt]
\multicolumn{6}{l}{\textit{Quark--gluon dataset}} \\
\addlinespace[2pt]
ParT\cite{Qu:2022mxj} & 0.849 & 0.9203 & $47.9\pm0.5$ & $129.5\pm0.9$ & 2.14M (100\%) \\
ParT-HG (this work) & 0.850 & 0.9206 & $48.9\pm0.4$ & $133.6\pm4.3$ & 1.03M (48.1\%) \\
MIParT\cite{Wu:2024thh} & 0.851 & 0.9215 & $49.3\pm0.4$ & $133.9\pm1.4$ & 657.9k (100\%) \\
MIParT-HG (this work) & 0.851 & 0.9216 & $48.7\pm0.3$ & $132.9\pm3$ & 261.3k (39.7\%) \\
\bottomrule
\end{tabular*}
\end{table*}

Table~\ref{tab:mipart_parameter_breakdown} gives the corresponding comparison for MIParT. The particle-embedding contribution decreases from 102.0k to 4.2k parameters, and the combined FFN contribution decreases from 405.5k to 106.8k. Together, the two modifications reduce the total parameter count from 657.9k to 261.3k, a reduction of 60.3\%.

We additionally define MIParT-L-HG as a larger MIParT configuration with $d=128$. It uses the same two modifications and the embedding path $C\to4C\to64\to128$. Its parameter count decreases from 2.38M to 985.6k, corresponding to a reduction of 58.6\%.

\section{Results and Discussions}
\label{sec:results}

All experiments are implemented in PyTorch\cite{Paszke:2019xhz} using the Weaver framework\cite{Weaver:software}. Unless otherwise stated, we use a learning rate of 0.001, a batch size of 256, and 20 training epochs on a single NVIDIA RTX 4090 GPU.

\subsection{Top and Quark--Gluon Tagging Performance}
\label{subsec:binary_benchmarks}

We evaluate ParT-HG and MIParT-HG on the top tagging dataset and the quark--gluon dataset. In the top tagging dataset, hadronically decaying top jets are the signal, while a mixture of quark and gluon jets forms the background\cite{Kasieczka:2019dbj,TopTagging:2019zenodo}. In the quark--gluon dataset, quark jets are the signal and gluon jets are the background\cite{QGJets:2019zenodo}.

\begin{figure*}[!t]
\centering
\includegraphics[width=\textwidth]{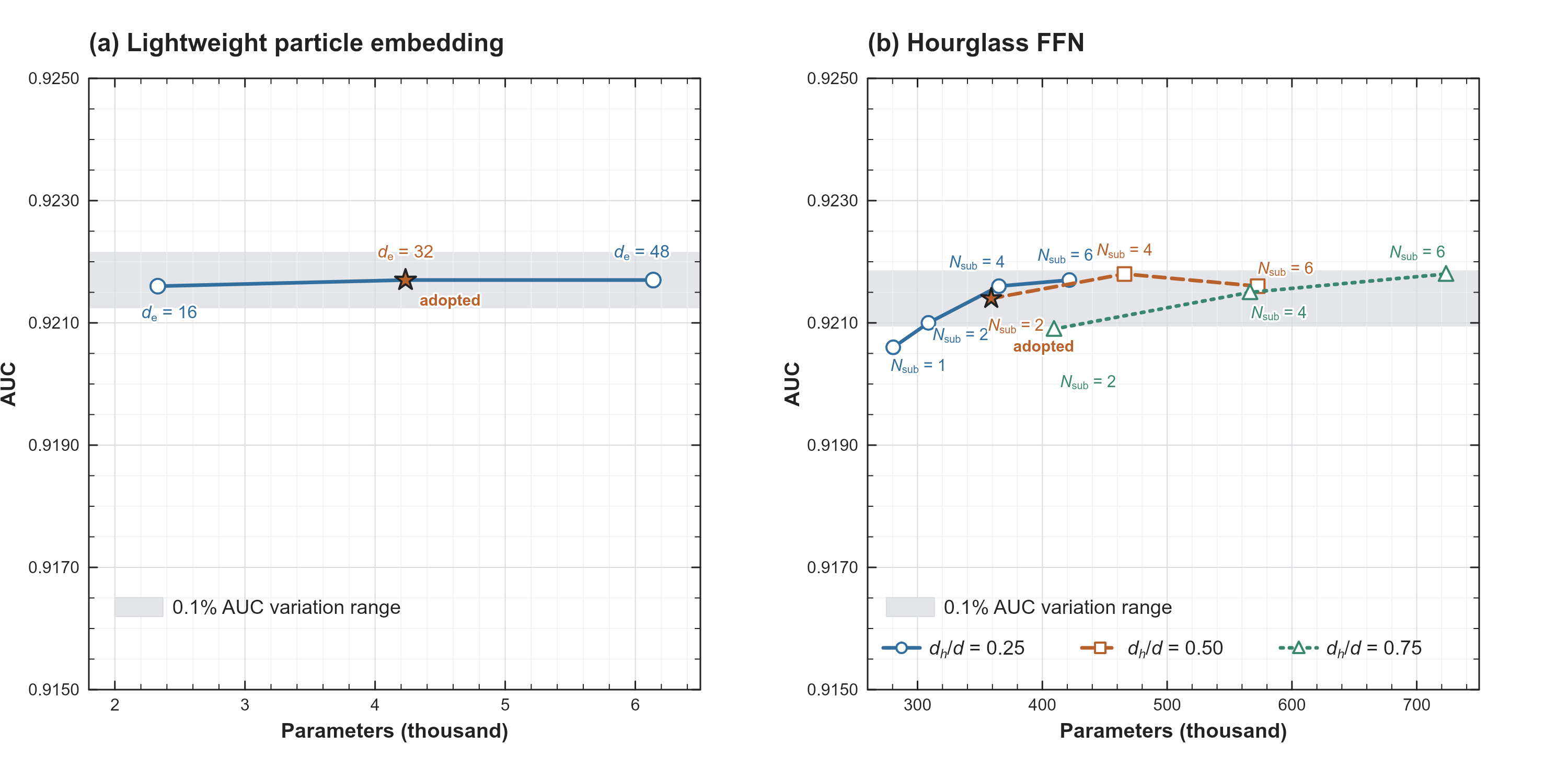}
\caption{Component ablations for MIParT on the quark--gluon dataset. 
The horizontal axes show the particle-embedding parameter count in (a) and the total trainable parameter count in (b), both in thousands. 
Point labels in (a) give $d_e$. 
In (b), colors, markers, and line styles denote the bottleneck ratio $d_h/d$, and point labels give $N_{\mathrm{sub}}$. 
Stars mark the adopted configurations. 
Each light-gray band spans 0.1\% of the adopted AUC value and is centered on that configuration.}
\label{fig:component_ablations}
\end{figure*}

We quantify performance at fixed signal efficiency using the background rejection $\mathrm{Rej}_{X}=1/\epsilon_B$, evaluated at $\epsilon_S=X$, where $\epsilon_S$ is the signal efficiency and $\epsilon_B$ is the background mistag efficiency. Thus, $\mathrm{Rej}_{50}$ and $\mathrm{Rej}_{30}$ correspond to signal efficiencies of 50\% and 30\%, respectively, and larger values indicate better background suppression.

Figure~\ref{fig:auc_vs_parameters} compares AUC and trainable parameter count before and after HG compression. The horizontal axis shows model size in trainable parameters on a linear scale, while the vertical axis shows the area under the receiver operating characteristic curve (AUC), for which higher values indicate better discrimination. Panels (a) and (b) show the top tagging and quark--gluon datasets, respectively. Gray points denote other published models, while arrows connect ParT to ParT-HG and MIParT to MIParT-HG. On the top tagging dataset, ParT-HG and MIParT-HG reduce the parameter counts of ParT and MIParT by 51.9\% and 60.3\%, respectively, while their AUCs remain nearly unchanged, moving from 0.9858 to 0.9861 and from 0.9868 to 0.9865 (a 0.03\% relative increase and a 0.03\% relative decrease). On the quark--gluon dataset, ParT-HG and MIParT-HG again reduce the corresponding baseline parameter counts by 51.9\% and 60.3\%, while their AUCs remain nearly unchanged, moving from 0.9203 to 0.9206 and from 0.9215 to 0.9216 (relative increases of 0.03\% and 0.01\%, respectively). In both panels, the dominant movement is leftward: parameter counts decrease substantially, whereas AUC values change only slightly. Table~\ref{tab:binary_performance} gives the full numerical results.

\subsection{Lightweight Particle Embedding and Hourglass FFN Ablations}
\label{subsec:ablation}

\begin{table*}[tbph]
\centering
\footnotesize
\setlength{\tabcolsep}{3pt}
\caption{Performance on JetClass (10M). Accuracy and AUC are aggregate metrics over all classes. Quark and gluon jets are treated as background for the class-specific rejection values. The signal efficiency is 50\% except for $H\to\ell\nu qq'$ and $t\to b\ell\nu$, where it is 99\% and 99.5\%, respectively. The baseline values are taken from the cited references, whereas the ParT-HG and MIParT-L-HG values are obtained in this work.}
\label{tab:jetclass_performance}

\begin{tabular*}{\textwidth}{
@{\extracolsep{\fill}}
l c c c c c c c c c c c c
@{}}
\toprule
& & \multicolumn{2}{c}{All classes}
& $H\to b\bar b$ & $H\to c\bar c$ & $H\to gg$ & $H\to 4q$ & $H\to \ell\nu qq'$ & $t\to bqq'$ & $t\to b\ell\nu$ & $W\to qq'$ & $Z\to qq'$ \\
Model & Parameters & Accuracy & AUC & $\mathrm{Rej}_{50}$ & $\mathrm{Rej}_{50}$ & $\mathrm{Rej}_{50}$ & $\mathrm{Rej}_{50}$ & $\mathrm{Rej}_{99}$ & $\mathrm{Rej}_{50}$ & $\mathrm{Rej}_{99.5}$ & $\mathrm{Rej}_{50}$ & $\mathrm{Rej}_{50}$ \\
\midrule
ParT~\cite{Qu:2022mxj} & 2.14M & 0.8500 & 0.9860 & 8734 & 3040 & 110 & 1274 & 3257 & 12579 & 8969 & 431 & 324 \\

ParT-HG & 1.03M & 0.8468 & 0.9856 & 8230 & 2766 & 109 & 1172 & 3306 & 12658 & 8584 & 417 & 311 \\

MIParT-L~\cite{Wu:2024thh} & 2.38M & 0.8500 & 0.9861 & 8000 & 3003 & 112 & 1281 & 3650 & 16529 & 9852 & 440 & 336 \\

MIParT-L-HG & 985.6k & 0.8470 & 0.9856 & 7843 & 2878 & 111 & 1214 & 3132 & 13072 & 8511 & 417 & 321 \\
\bottomrule
\end{tabular*}
\end{table*}

We perform both ablation studies with MIParT on the quark--gluon dataset. To isolate the lightweight particle embedding, we retain the standard MIParT FFNs and vary only the intermediate pointwise width $d_e$ in the mapping $4C\to d_e\to d$. We test $d_e/d=0.25$, $0.50$, and $0.75$, corresponding to $d_e=16$, 32, and 48 for $d=64$. To isolate the hourglass FFN, we retain the original dense particle embedding. For $d_h/d=0.25$, we test $N_{\mathrm{sub}}=1$, 2, 4, and 6; for $d_h/d=0.50$ and 0.75, we test $N_{\mathrm{sub}}=2$, 4, and 6.

In the left plot of Fig.~\ref{fig:component_ablations}, the horizontal axis gives the trainable parameter count of the particle embedding, and the vertical axis gives AUC. Increasing $d_e$ from 16 to 32 and 48 raises the embedding parameter count from 2.330k to 4.234k and 6.138k, while the AUC changes only from 0.9216 to 0.9217. We therefore adopt the intermediate ratio $d_e/d=0.50$, rather than either endpoint, as a representative setting for the HG models.
In the right plot, the horizontal and vertical axes again give the total trainable parameter count and AUC, respectively. The three colored curves correspond to $d_h/d=0.25$, 0.50, and 0.75, while the integer beside each point gives $N_{\mathrm{sub}}$. All three ratios include $N_{\mathrm{sub}}=2$, 4, and 6, and the $d_h/d=0.25$ curve also includes $N_{\mathrm{sub}}=1$. At fixed $d_h/d$, increasing $N_{\mathrm{sub}}$ generally raises the AUC, although the trend is not strictly monotonic, and also increases the parameter count. At fixed depth, increasing $d_h/d$ likewise increases the parameter count. We therefore adopt the intermediate configuration $(N_{\mathrm{sub}},d_h/d)=(2,0.50)$ as a representative choice that avoids both the smallest and largest tested models.

\subsection{Larger-Scale Evaluation}
\label{subsec:jetclass}

To investigate the effect of parameter compression on a larger-scale dataset, we compare ParT-HG and \mbox{MIParT-L-HG} with ParT and MIParT-L on JetClass. The JetClass dataset focuses on identifying jets originating from Lorentz-boosted $W$ and $Z$ bosons, Higgs bosons, and top quarks\cite{Qu:2022mxj,JetClass:2022zenodo}. We use 10 million jets for training. MIParT-L increases the MIParT embedding dimension from $d=64$ to $d=128$\cite{Wu:2024thh}, while \mbox{MIParT-L-HG} applies the compressed particle embedding and FFNs to this larger baseline. For the JetClass evaluation, both ParT-HG and MIParT-L-HG use a batch size of 512 and 50 training epochs.

Table~\ref{tab:jetclass_performance} shows that ParT-HG and \mbox{MIParT-L-HG} reduce their baseline parameter counts by 51.9\% and 58.6\%, while their accuracies decrease by 0.32\% and 0.30\% and their AUCs decrease by 0.04\% and 0.05\%, respectively. Most class-specific background-rejection values decrease for ParT-HG, most strongly by 9.0\% for $H\to c\bar c$ and 8.0\% for $H\to4q$. All reported values decrease for \mbox{MIParT-L-HG}, with the largest reductions of 20.9\% for $t\to bqq'$, 14.2\% for $H\to\ell\nu qq'$, and 13.6\% for $t\to b\ell\nu$. Because $\mathrm{Rej}_{X}=1/\epsilon_B$, small absolute changes in the low background mistag efficiency can produce much larger numerical changes in rejection. The JetClass results therefore make the performance cost of compression more visible and suggest that larger datasets may require greater model capacity.


\section{Conclusion}
\label{sec:conclusion}

We investigated whether Transformer-based jet taggers contain parameters that can be removed while retaining tagging performance. In ParT and MIParT, the standard FFNs contain most of the parameters, and the dense particle embedding provides another target for reduction. We replaced these components with residual hourglass FFNs and a lightweight particle embedding to obtain ParT-HG and MIParT-HG. We evaluated both models on the top tagging and quark--gluon datasets and performed separate ablations of the two designs. We also applied the method to the larger multiclass JetClass dataset.

On the top tagging and quark--gluon datasets, ParT-HG and MIParT-HG reduce the parameter counts of their baselines by 51.9\% and 60.3\%, respectively, while leaving tagging performance almost unchanged. On JetClass, accuracy and AUC decrease by less than 1\%, and background rejection also decreases for several signal classes. These results show that substantial parameter reduction is possible on the standard benchmarks, whereas a much larger training sample can expose a loss in background suppression.

This scale dependence provides a practical design principle for jet-tagging analyses. For example, the training sample may be limited when full detector simulation is costly. In this situation, the full Transformer may be unnecessarily large and contain more parameters than the data can constrain. Our method reduces the FFN and particle-embedding parameters while retaining the particle-interaction attention that encodes correlations among jet constituents. Our work therefore provides an alternative approach to reducing model parameters while retaining jet-tagging performance.

\acknowledgments
We thank Xinzhu Wang and Chunxiang Zhu for useful discussions and comments. 
This work was supported by the National Natural Science Foundation of China under Grant No. 12275066 and by the startup research funds of Henan University. 
The work of H. Yang was also supported by the National Natural Science Foundation of China under Grant number: W2441004.
And the work of K. Wang was also supported by the Open Project of the Shanghai Key Laboratory of Particle Physics and Cosmology under Grant No. 22DZ2229013-3.


\section*{DATA AVAILABILITY}

The source code used in this work is publicly available at
\url{https://github.com/Cforik-L/ParT-MIParT-HG}.

\bibliographystyle{apsrev4-1}
\bibliography{apssamp}

@article{Larkoski:2017jix,
  author        = {Larkoski, Andrew J. and Moult, Ian and Nachman, Benjamin},
  title         = {Jet Substructure at the Large Hadron Collider: A Review of Recent Advances in Theory and Machine Learning},
  eprint        = {1709.04464},
  archivePrefix = {arXiv},
  primaryClass  = {hep-ph},
  doi           = {10.1016/j.physrep.2019.11.001},
  journal       = {Phys. Rept.},
  volume        = {841},
  pages         = {1--63},
  year          = {2020}
}

@article{Kogler:2018hem,
  author        = {Kogler, Roman and others},
  title         = {Jet Substructure at the Large Hadron Collider: Experimental Review},
  eprint        = {1803.06991},
  archivePrefix = {arXiv},
  primaryClass  = {hep-ex},
  doi           = {10.1103/RevModPhys.91.045003},
  journal       = {Rev. Mod. Phys.},
  volume        = {91},
  number        = {4},
  pages         = {045003},
  year          = {2019}
}

@article{Thaler:2010tr,
  author        = {Thaler, Jesse and Van Tilburg, Ken},
  title         = {Identifying Boosted Objects with N-subjettiness},
  eprint        = {1011.2268},
  archivePrefix = {arXiv},
  primaryClass  = {hep-ph},
  doi           = {10.1007/JHEP03(2011)015},
  journal       = {JHEP},
  volume        = {03},
  pages         = {015},
  year          = {2011}
}

@article{Larkoski:2013eya,
  author        = {Larkoski, Andrew J. and Salam, Gavin P. and Thaler, Jesse},
  title         = {Energy Correlation Functions for Jet Substructure},
  eprint        = {1305.0007},
  archivePrefix = {arXiv},
  primaryClass  = {hep-ph},
  doi           = {10.1007/JHEP06(2013)108},
  journal       = {JHEP},
  volume        = {06},
  pages         = {108},
  year          = {2013}
}

@article{Larkoski:2014gra,
  author        = {Larkoski, Andrew J. and Moult, Ian and Neill, Duff},
  title         = {Power Counting to Better Jet Observables},
  eprint        = {1409.6298},
  archivePrefix = {arXiv},
  primaryClass  = {hep-ph},
  doi           = {10.1007/JHEP12(2014)009},
  journal       = {JHEP},
  volume        = {12},
  pages         = {009},
  year          = {2014}
}

@article{Dasgupta:2013ihk,
  author        = {Dasgupta, Mrinal and Fregoso, Alessandro and Marzani, Simone and Salam, Gavin P.},
  title         = {Towards an Understanding of Jet Substructure},
  eprint        = {1307.0007},
  archivePrefix = {arXiv},
  primaryClass  = {hep-ph},
  doi           = {10.1007/JHEP09(2013)029},
  journal       = {JHEP},
  volume        = {09},
  pages         = {029},
  year          = {2013}
}

@article{Larkoski:2014wba,
  author        = {Larkoski, Andrew J. and Marzani, Simone and Soyez, Gregory and Thaler, Jesse},
  title         = {Soft Drop},
  eprint        = {1402.2657},
  archivePrefix = {arXiv},
  primaryClass  = {hep-ph},
  doi           = {10.1007/JHEP05(2014)146},
  journal       = {JHEP},
  volume        = {05},
  pages         = {146},
  year          = {2014}
}

@article{Krohn:2012fg,
  author        = {Krohn, David and Schwartz, Matthew D. and Lin, Tongyan and Waalewijn, Wouter J.},
  title         = {Jet Charge at the LHC},
  eprint        = {1209.2421},
  archivePrefix = {arXiv},
  primaryClass  = {hep-ph},
  doi           = {10.1103/PhysRevLett.110.212001},
  journal       = {Phys. Rev. Lett.},
  volume        = {110},
  number        = {21},
  pages         = {212001},
  year          = {2013}
}

@article{Cogan:2014oua,
  author        = {Cogan, Josh and Kagan, Michael and Strauss, Emanuel and Schwartzman, Ariel},
  title         = {Jet-Images: Computer Vision Inspired Techniques for Jet Tagging},
  eprint        = {1407.5675},
  archivePrefix = {arXiv},
  primaryClass  = {hep-ph},
  doi           = {10.1007/JHEP02(2015)118},
  journal       = {JHEP},
  volume        = {02},
  pages         = {118},
  year          = {2015}
}

@article{deOliveira:2015xxd,
  author        = {de Oliveira, Luke and Kagan, Michael and Mackey, Lester and Nachman, Benjamin and Schwartzman, Ariel},
  title         = {Jet-images---deep learning edition},
  eprint        = {1511.05190},
  archivePrefix = {arXiv},
  primaryClass  = {hep-ph},
  doi           = {10.1007/JHEP07(2016)069},
  journal       = {JHEP},
  volume        = {07},
  pages         = {069},
  year          = {2016}
}

@inproceedings{Zaheer:2017wmg,
  author        = {Zaheer, Manzil and Kottur, Satwik and Ravanbakhsh, Siamak and Poczos, Barnabas and Salakhutdinov, Ruslan and Smola, Alexander J.},
  title         = {Deep Sets},
  eprint        = {1703.06114},
  archivePrefix = {arXiv},
  primaryClass  = {cs.LG},
  booktitle     = {Advances in Neural Information Processing Systems 30},
  year          = {2017}
}

@article{Komiske:2018cqr,
  author        = {Komiske, Patrick T. and Metodiev, Eric M. and Thaler, Jesse},
  title         = {Energy Flow Networks: Deep Sets for Particle Jets},
  eprint        = {1810.05165},
  archivePrefix = {arXiv},
  primaryClass  = {hep-ph},
  doi           = {10.1007/JHEP01(2019)121},
  journal       = {JHEP},
  volume        = {01},
  pages         = {121},
  year          = {2019}
}

@article{Qu:2019gqs,
  author        = {Qu, Huilin and Gouskos, Loukas},
  title         = {ParticleNet: Jet Tagging via Particle Clouds},
  eprint        = {1902.08570},
  archivePrefix = {arXiv},
  primaryClass  = {hep-ph},
  doi           = {10.1103/PhysRevD.101.056019},
  journal       = {Phys. Rev. D},
  volume        = {101},
  number        = {5},
  pages         = {056019},
  year          = {2020}
}

@article{Gong:2022lye,
  author        = {Gong, Shiqi and Meng, Qi and Zhang, Jue and Qu, Huilin and Li, Congqiao and Qian, Sitian and Du, Weitao and Ma, Zhi-Ming and Liu, Tie-Yan},
  title         = {An Efficient Lorentz Equivariant Graph Neural Network for Jet Tagging},
  eprint        = {2201.08187},
  archivePrefix = {arXiv},
  primaryClass  = {hep-ph},
  doi           = {10.1007/JHEP07(2022)030},
  journal       = {JHEP},
  volume        = {07},
  pages         = {030},
  year          = {2022}
}

@article{Bogatskiy:2023nnw,
  author        = {Bogatskiy, Alexander and Hoffman, Timothy and Miller, David W. and Offermann, Jan T. and Liu, Xiaoyang},
  title         = {Explainable Equivariant Neural Networks for Particle Physics: PELICAN},
  eprint        = {2307.16506},
  archivePrefix = {arXiv},
  primaryClass  = {hep-ph},
  doi           = {10.1007/JHEP03(2024)113},
  journal       = {JHEP},
  volume        = {03},
  pages         = {113},
  year          = {2024}
}

@article{Brehmer:2024yqw,
  author        = {Brehmer, Johann and Bres{\'o}, V{\'i}ctor and de Haan, Pim and Plehn, Tilman and Qu, Huilin and Spinner, Jonas and Thaler, Jesse},
  title         = {A Lorentz-Equivariant Transformer for All of the LHC},
  eprint        = {2411.00446},
  archivePrefix = {arXiv},
  primaryClass  = {hep-ph},
  doi           = {10.21468/SciPostPhys.19.4.108},
  journal       = {SciPost Phys.},
  volume        = {19},
  number        = {4},
  pages         = {108},
  year          = {2025}
}

@inproceedings{Vaswani:2017lxt,
  author        = {Vaswani, Ashish and Shazeer, Noam and Parmar, Niki and Uszkoreit, Jakob and Jones, Llion and Gomez, Aidan N. and Kaiser, Lukasz and Polosukhin, Illia},
  title         = {Attention Is All You Need},
  eprint        = {1706.03762},
  archivePrefix = {arXiv},
  primaryClass  = {cs.CL},
  booktitle     = {Advances in Neural Information Processing Systems 30},
  year          = {2017}
}

@inproceedings{Qu:2022mxj,
  author        = {Qu, Huilin and Li, Congqiao and Qian, Sitian},
  title         = {Particle Transformer for Jet Tagging},
  eprint        = {2202.03772},
  archivePrefix = {arXiv},
  primaryClass  = {hep-ph},
  booktitle     = {Proceedings of the 39th International Conference on Machine Learning},
  series        = {Proceedings of Machine Learning Research},
  volume        = {162},
  pages         = {18281--18292},
  publisher     = {PMLR},
  url           = {https://proceedings.mlr.press/v162/qu22b.html},
  year          = {2022}
}

@article{Wu:2024thh,
  author        = {Wu, Yifan and Wang, Kun and Li, Congqiao and Qu, Huilin and Zhu, Jingya},
  title         = {Jet Tagging with More-Interaction Particle Transformer},
  eprint        = {2407.08682},
  archivePrefix = {arXiv},
  primaryClass  = {hep-ph},
  doi           = {10.1088/1674-1137/ad7f3d},
  journal       = {Chin. Phys. C},
  volume        = {49},
  number        = {1},
  pages         = {013110},
  year          = {2025}
}

@article{Esmail:2025kii,
  author        = {Esmail, Waleed and Hammad, Ahmed and Nojiri, Mihoko M.},
  title         = {IAFormer: Interaction-Aware Transformer Network for Collider Data Analysis},
  eprint        = {2505.03258},
  archivePrefix = {arXiv},
  primaryClass  = {hep-ph},
  doi           = {10.21468/SciPostPhys.20.4.108},
  journal       = {SciPost Phys.},
  volume        = {20},
  number        = {4},
  pages         = {108},
  year          = {2026}
}

@article{Rai:2025bitpart,
  author        = {Rai, Saurabh and {Prisha} and Kumar, Jitendra},
  title         = {Investigating 1-Bit Quantization in Transformer-Based Top Tagging},
  eprint        = {2508.07431},
  archivePrefix = {arXiv},
  primaryClass  = {hep-ph},
  year          = {2025},
  note          = {Preprint}
}

@article{Wang:2025dky,
  author        = {Wang, Aaron and Zhao, Zihan and Katel, Subash and Sahu, Vivekanand Gyanchand and Khoda, Elham E. and Gandrakota, Abhijith and Ngadiuba, Jennifer and Cavanaugh, Richard and Duarte, Javier},
  title         = {Spatially Aware Linear Transformer (SAL-T) for Particle Jet Tagging},
  eprint        = {2510.23641},
  archivePrefix = {arXiv},
  primaryClass  = {cs.LG},
  year          = {2025},
  note          = {Preprint}
}

@article{Zheng:2026juf,
  author        = {Zheng, Ruoqing and Sun, Chang and Liu, Qibin and Laatu, Lauri and Cox, Arianna and Maier, Benedikt and Tapper, Alexander and Coutinho, Jose G. F. and Luk, Wayne and Que, Zhiqiang},
  title         = {JetFormer: A Scalable and Efficient Transformer for Jet Tagging from Offline Analysis to FPGA Triggers},
  eprint        = {2601.17215},
  archivePrefix = {arXiv},
  primaryClass  = {cs.LG},
  year          = {2026},
  note          = {Preprint}
}

@article{Wang:2026lqb,
  author        = {Wang, Aaron and Zhao, Zihan and Xia, Alan and Sun, Chang and Gandrakota, Abhijith and Ngadiuba, Jennifer and Cavanaugh, Richard and Duarte, Javier},
  title         = {Patch Hierarchical Attention Transformer for Efficient Particle Jet Tagging},
  eprint        = {2605.21789},
  archivePrefix = {arXiv},
  primaryClass  = {hep-ex},
  year          = {2026},
  note          = {Preprint}
}

@inproceedings{Jiang:2024lvg,
  author        = {Jiang, Zhixing and Yin, Dennis and Khoda, Elham E. and Loncar, Vladimir and Govorkova, Ekaterina and Moreno, Eric and Harris, Philip and Hauck, Scott and Hsu, Shih-Chieh},
  title         = {Ultra Fast Transformers on {FPGA}s for Particle Physics Experiments},
  booktitle     = {Machine Learning and the Physical Sciences Workshop, NeurIPS 2023},
  eprint        = {2402.01047},
  archivePrefix = {arXiv},
  primaryClass  = {cs.LG},
  year          = {2023}
}

@article{Laatu:2025xsw,
  author        = {Laatu, Lauri and Sun, Chang and Cox, Arianna and Gandrakota, Abhijith and Maier, Benedikt and Ngadiuba, Jennifer and Que, Zhiqiang and Luk, Wayne and Spiropulu, Maria and Tapper, Alexander},
  title         = {Sub-microsecond Transformers for Jet Tagging on FPGAs},
  eprint        = {2510.24784},
  archivePrefix = {arXiv},
  primaryClass  = {physics.ins-det},
  year          = {2025},
  note          = {Preprint}
}

@inproceedings{Koski:2026rcc,
  author        = {Koski, Gram and Lipps, Sean and Ma, Zhenghua and Abarajithan, G. and Kastner, Ryan},
  title         = {Reconfigurable Computing Challenge: Transformer for Jet Tagging on Versal AI Engines},
  eprint        = {2606.17500},
  archivePrefix = {arXiv},
  primaryClass  = {cs.LG},
  doi           = {10.1109/FCCM68464.2026.00078},
  booktitle     = {2026 IEEE 34th International Symposium on Field-Programmable Custom Computing Machines},
  pages         = {307--310},
  year          = {2026},
}

@article{Petitjean:2025zjf,
  author        = {Petitjean, Antoine and Plehn, Tilman and Spinner, Jonas and K{\"o}the, Ullrich},
  title         = {Economical Jet Taggers---Equivariant, Slim, and Quantized},
  eprint        = {2512.17011},
  archivePrefix = {arXiv},
  primaryClass  = {hep-ph},
  reportNumber  = {IPPP/25/93},
  month         = {12},
  year          = {2025},
  note          = {Preprint}
}

@article{Liao:2026hourglass,
  author        = {Liao, Feng-Ting and Chen, Meng-Hsi and Yi, Guan-Ting and Shiu, Da-shan},
  title         = {Revisiting the Shape Convention of Transformer Language Models},
  eprint        = {2602.06471},
  archivePrefix = {arXiv},
  primaryClass  = {cs.CL},
  year          = {2026},
  note          = {Preprint}
}

@article{Chen:2025hourglass,
  author        = {Chen, Meng-Hsi and Lee, Yu-Ang and Liao, Feng-Ting and Shiu, Da-shan},
  title         = {Rethinking the Shape Convention of an MLP},
  eprint        = {2510.01796},
  archivePrefix = {arXiv},
  primaryClass  = {cs.LG},
  year          = {2025},
  note          = {Preprint}
}

@inproceedings{He:2015wrn,
  author        = {He, Kaiming and Zhang, Xiangyu and Ren, Shaoqing and Sun, Jian},
  title         = {Deep Residual Learning for Image Recognition},
  eprint        = {1512.03385},
  archivePrefix = {arXiv},
  primaryClass  = {cs.CV},
  doi           = {10.1109/CVPR.2016.90},
  booktitle     = {2016 IEEE Conference on Computer Vision and Pattern Recognition},
  pages         = {770--778},
  year          = {2016}
}

@inproceedings{Chollet:2016xception,
  author        = {Chollet, Fran{\c{c}}ois},
  title         = {Xception: Deep Learning with Depthwise Separable Convolutions},
  eprint        = {1610.02357},
  archivePrefix = {arXiv},
  primaryClass  = {cs.CV},
  doi           = {10.1109/CVPR.2017.195},
  booktitle     = {2017 IEEE Conference on Computer Vision and Pattern Recognition},
  pages         = {1800--1807},
  year          = {2017}
}

@inproceedings{Sandler:2018mobilenetv2,
  author        = {Sandler, Mark and Howard, Andrew and Zhu, Menglong and Zhmoginov, Andrey and Chen, Liang-Chieh},
  title         = {MobileNetV2: Inverted Residuals and Linear Bottlenecks},
  eprint        = {1801.04381},
  archivePrefix = {arXiv},
  primaryClass  = {cs.CV},
  doi           = {10.1109/CVPR.2018.00474},
  booktitle     = {2018 IEEE/CVF Conference on Computer Vision and Pattern Recognition},
  pages         = {4510--4520},
  year          = {2018}
}

@article{Kasieczka:2019dbj,
  author = {Kasieczka, Gregor and others},
  title         = {The Machine Learning Landscape of Top Taggers},
  eprint        = {1902.09914},
  archivePrefix = {arXiv},
  primaryClass  = {hep-ph},
  doi           = {10.21468/SciPostPhys.7.1.014},
  journal       = {SciPost Phys.},
  volume        = {7},
  number        = {1},
  pages         = {014},
  year          = {2019}
}

@misc{TopTagging:2019zenodo,
  author    = {Kasieczka, Gregor and Plehn, Tilman and Thompson, Jennifer and Russel, Michael},
  title     = {Top Quark Tagging Reference Dataset},
  publisher = {Zenodo},
  version   = {v0 (2018_03_27)},
  doi       = {10.5281/zenodo.2603256},
  url       = {https://doi.org/10.5281/zenodo.2603256},
  year      = {2019}
}

@misc{QGJets:2019zenodo,
  author    = {Komiske, Patrick T. and Metodiev, Eric M. and Thaler, Jesse},
  title     = {{Pythia8} Quark and Gluon Jets for Energy Flow},
  howpublished = {\url{https://doi.org/10.5281/zenodo.3164691}},
  publisher = {Zenodo},
  version   = {v1},
  doi       = {10.5281/zenodo.3164691},
  url       = {https://doi.org/10.5281/zenodo.3164691},
  year      = {2019}
}

@misc{JetClass:2022zenodo,
  author    = {Qu, Huilin and Li, Congqiao and Qian, Sitian},
  title     = {{JetClass}: A Large-Scale Dataset for Deep Learning in Jet Physics},
  publisher = {Zenodo},
  version   = {1.0.0},
  doi       = {10.5281/zenodo.6619768},
  howpublished = {\url{https://doi.org/10.5281/zenodo.6619768}},
  url       = {https://doi.org/10.5281/zenodo.6619768},
  year      = {2022}
}

@inproceedings{Paszke:2019xhz,
  author        = {Paszke, Adam and others},
  title         = {PyTorch: An Imperative Style, High-Performance Deep Learning Library},
  eprint        = {1912.01703},
  archivePrefix = {arXiv},
  primaryClass  = {cs.LG},
  booktitle     = {Advances in Neural Information Processing Systems 32},
  year          = {2019}
}

@misc{Weaver:software,
  author       = {Qu, Huilin},
  title        = {Weaver: Streamlined Neural Network Training for High Energy Physics},
  howpublished = {\url{https://github.com/hqucms/weaver-core}},
  version      = {0.5.3},
  year         = {2026},
  note         = {Software repository; accessed 2026-07-24}
}

\end{document}